\documentclass[prb,twocolumn]{revtex4}
\usepackage{amssymb}
\usepackage{graphicx}
\usepackage{dsfont}
\def\braket#1{\left(#1\right)}
\def\sb#1{\left[#1\right]}

\def\ket#1{\left|#1\right\rangle}

\def\'{\prime}

\begin{document}
\title{Grassmann tensor network states and its renormalization\\
for strongly correlated fermionic and bosonic states}
\author{Zheng-Cheng Gu$^\dagger$, Frank Verstraete$^{\dagger\dagger}$ and  Xiao-Gang Wen$^{\dagger\dagger\dagger}$}
\affiliation{Kavli Institute for Theoretical Physics, University of
California, Santa Barbara, CA 93106, USA$^{\dagger}$ \\
 Fakult\"{a}t f\"{u}r Physik, Universit\"{a}t Wien, Boltzmanngasse 5, A-1090 Wien
 $^{\dagger\dagger}$\\
 Department
of Physics, Massachusetts Institute of Technology, Cambridge,
Massachusetts 02139, USA$^{\dagger\dagger\dagger}$}

\begin{abstract}
The projective construction (the slave-particle approach) has played
an very important role in understanding strongly correlated systems,
such as the emergence of fermions, anyons, and gauge theory in
quantum spin liquids and quantum Hall states.  Recently, fermionic
Projected Entangled Pair States (fPEPS) have  been introduced to
efficiently represent many-body fermionic states.  In this paper, we
show that the strongly correlated bosonic/fermionic states obtained
both from the projective construction and the fPEPS approach can be
represented systematically as Grassmann tensor product states. This
construction can also be applied to all other tensor network states
approaches. The Grassmann tensor product states allow us to encode
many-body bosonic/fermionic states efficiently with a polynomial
number of parameters.  We also generalize the tensor-entanglement
renormalization group (TERG) method for complex tensor networks to
Grassmann tensor networks. This allows us to approximate the norm
and average local operators of Grassmann tensor product states in
polynomial time, and hence leads to a variational approach for
describing strongly correlated bosonic/fermionic systems in higher
dimensions.

\end{abstract}
\maketitle

\section{Introduction}

Traditional condensed matter physics is based on two theories:
symmetry breaking theory for phases and phase transitions, and Fermi
liquid theory for metals. Within the Fermi liquid theory, one assumes
that the ground state wave function for the electrons can be
approximately described by a Slater determinant. In other words, one
assumes that the many-electron ground state can be constructed by
filling the single-particle energy levels. Such an energy-level
filling picture becomes a foundation for traditional many-body
physics.  In this paper, we will call states obtained by filling
single-particle energy levels as energy-level-filling (ELF) states.
In addition to Slater determinant states, fermion paired states are
also ELF states.

We may view the ELF construction of many-body states as an
encoding method of a physically relevant subset of states.   Although a random many-body state can only specified by an exponential amount of data, hence making it impossible to specify and calculate physical properties efficiently, physically relevant states seem to have a much simpler entanglement structure. A generic ELF state on a lattice
can be written as
\begin{align}
 |\Psi_f\>= \exp\Big(\sum_{\<ij\>} u_{ij} c^\dag_j c^\dag_i \Big)|0\>
\end{align}
where $\sum_{\<ij\>}$ sums over the pairs of sites in the
lattice.  Here we consider a simple example of paired
spinless fermions, and unpaired fermions can also be represented as limits of such states. Such a many-body fermionic state is encoded
by polynomial amount information characterized by $u_{ij}$. A
crucial property is that it is very easy to calculate
the norm and the averages of any local operators  for a ELF
state;  here ``easy'' means that the computational cost scales as a polynomial in the number of modes.
This effective encoding and the ease of calculating physical
quantities (such as energy) form the foundation of the
standard mean-field theory for interacting electron systems.
In such a ELF approach, we may view $u_{ij}$ as variational
parameters and minimize the average energy by varying
$u_{ij}$.  The $\bar u_{ij}$ that minimize the energy give
us an approximated many-fermion ground state.  From the form
of $\bar u_{ij}$ we can determine which symmetry is broken
and obtain the phase diagram of the system.

However, the ground states for some strongly correlated electron
systems cannot be approximated by ELF states, i.e. they cannot be constructed by filling energy levels.  One classic
example is the filling fraction $\nu=1/3$ Laughlin state\cite{L8395}
\begin{equation}
 \Psi_3=\prod_{i<j}(z_i-z_j)^3 \e^{-\frac14 \sum_i |z_i|^2}
\end{equation}
Although the $\nu=1$ state $\Psi_1=\prod_{i<j}(z_i-z_j) \e^{-\frac14
\sum_i |z_i|^2}$ is a ELF state, its cubic power, $\Psi_3\sim
(\Psi_1)^3$, is very different from any ELF states.  In order to
obtain the low energy effective theories for systems that cannot be
described by ELF picture such as spin liquids and non-Fermi-liquid
metallic states, a projective construction (also known as
slave-particle approach or parton construction; see appendix
\ref{pcon} for a brief introduction) was
developed\cite{BZA8773,BA8880,AM8874,AZH8845,DFM8826,WWZcsp,Wsrvb,Wnab,Wpcon,Wen04,Wqoslpub}.
Those states may appear in high $T_c$ superconductors and other
strongly interacting systems, and are by now widely used. Just like
the ELF states, the projective states can also be characterized by a
polynomial amount of information. So the projective approach can
also be viewed as an efficient encoding of many-body states.

In the projective approach, we view the projective states as variational trial wave
functions\cite{Wqoslpub,Wen04} for  obtaining approximate ground state.  What
makes the projective approach so attractive is that from the form of the projective
state, we can usually obtain the low energy effective theory
that describes low energy excitations\cite{Wpcon,Wqoslpub}.
From the low energy effective theory, we have learned that projective states can
capture many qualitatively new phenomena that ELF states
fail to describe, such as fractional charge, fractional
statistics and topological orders\cite{WWZcsp,Wsrvb,Wpcon}.

However, it is much harder to calculate the norm and local expectation values in the projective approach.
 Although expectation values can efficiently be calculated using variational Monte Carlo methods
 in the case that electron or spin operators can be
expressed as the products of parton operators \cite{G8953,YS8882},
Monte Carlo fails when those operators are expressed as sums of the
products of parton operators \cite{Wpcon,RWffmf,RW0634}. As such
there is no general efficient way to calculate the norm and the
average of local operators.

In a different development, concepts of quantum information theory
have allowed one to gain a better insight into the entanglement
structure in ground states of generic Hamiltonians of strongly
correlated quantum spins. It has  been shown that ground states of
local Hamiltonians obey so-called area laws, and that ground states
can therefore be efficiently be represented by the class of
so-called Matrix Product States in one dimension or Projected
Entangled Pair States (PEPS) in higher dimensions
\cite{reviewMPS,VC0466}. This class of states share some
resemblances with the states obtained in the projective approach, as
they can be understood as projections of locally maximally entangled
pairs. Importantly, techniques have been developed that make it
possible to efficiently calculate expectation values of local
operators for this class of states, as this can be done by
contracting networks of tensor products.  In this paper, we will
show how to generalize such networks to include Grassmann tensors,
and show how they can be contracted efficiently.  This allows us to
define a large and important subclass of the states obtained in the
projective approach for which it is possible to calculate
expectation values. It also allows us to generalize all tensor
product state methods to the fermionic case; special subclasses
include the recently introduced fermionic PEPS (fPEPS)
\cite{FrankfPEPS} and the fermionic Multiscale Entanglement
Renormalization Ansatz \cite{fermionicMERA}.

The Grassmann tensor product states allow us to express the norms of
the projective wave functions in terms of Grassmann valued tensor
network. Similarly, the average of any local operators for the
projective wave functions can also be expressed in terms of
Grassmann tensor networks with a few ``impurity'' tensors.

Throughout our study, we will also find that the Grassmann tensor
product states can be more general than projective states.  Using
Grassman tensor networks, we can systematically construct
 more general strongly correlated states for both
bosonic and fermionic systems. We only need polynomial amount of
data to characterize the tensor network. Thus the Grassmann tensor
network approach gives us an effective encoding for both fermionic
and bosonic many-body states.

Calculating the norm and the expectation value of local operators
for a tensor product state can be exponentially hard in general
\cite{Schuch}.  However, many possible polynomial approximation
schemes, including the tensor-entanglement renormalization group
(TERG) method,  have been proposed in recent
years\cite{LN0701,JOV0802,GLWtergV,XiangTRG,GPEPS,finitefPEPS}. In
this paper, we will show how the TERG method\cite{GLWtergV} can also
be applied to Grassmann tensor network. In particular, this implies
that the average energy and other physical quantities of a Grassmann
tensor network state in two dimensions can be calculated efficiently
by using tensor network renormalization approach. This Grassmann
tensor network approach is the natural approach for expressing
fermions in tensor network methods and hence provides a new starting
point for studying strongly correlated bosonic/fermionic systems.

\section{Tensor-network representation of
ELF states}\label{ELF}

\begin{figure}
\begin{center}
\includegraphics[scale=0.5]{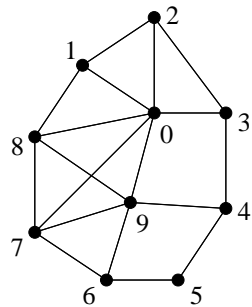}
\end{center}
\caption{
\label{rlatt}
A random lattice where the physical degrees of freedom are
localized on the vertices.
}
\end{figure}

Let us first consider a simple example of using tensor networks to
represent a ELF state on an
arbitrary graph, where the fermions live on the
vertices of the graph (see Fig. \ref{rlatt}):
\begin{align}
|\Psi_f \rangle=\exp\sb{\sum_{ ij } u_{ij}c_j^\dagger c_i^\dagger }
\ket{0}= \prod_{ij }\braket{1+ u_{ij}c_j^\dagger c_i^\dagger}
\ket{0}
\end{align}
As  $ u_{ij}c_j^\dagger c_i^\dagger =-u_{ij}c_i^\dagger
c_j^\dagger $, w.l.o.g. $u_{ij}=-u_{ji}$.  The many-body wave function $\Psi_f(\{m_i\})$ is given
by
\begin{equation}
\label{wvfree} \Psi_f(\{m_i\})= \<0| \prod_i (c_i)^{m_i} \prod_{ ij
}\braket{1+ u_{ij}c_j^\dagger c_i^\dagger} \ket{0}
\end{equation}
where $m_i=0,1$ indicating if the site-$i$ is empty or occupied.
Note that the fermion operators $c_i$ in  the product $\prod_i (c_i)^{m_i}$
is ordered in the following way
\begin{equation}
 \prod_i (c_i)^{m_i}\equiv
(c_1)^{m_1}
(c_2)^{m_2}
(c_3)^{m_3}
...
\end{equation}

Motivated by fermionic path integral, we may represent the above
wave function in terms of Grassmann numbers $\th_i$ and their
derivatives $\dd{\th_i}$, which satisfy
\begin{align}
 \th_i\th_j&=-\th_j\th_i,
&
 \dd{\th_i}\dd{\th_j}&=-\dd{\th_j}\dd{\th_i},
\nonumber\\
\int \dd{\th_i}\th_j &=\del_{ij}
&
\int \dd{\th_i} 1&=0 .
\end{align}
We find that the wave function can be
rewritten as
\begin{align}
\Psi_f(\{m_i\})&=\int \prod_i T^{m_i}_i \prod_{ij} G_{ij} ,
\nonumber\\
T^1_i &=\dd{\th_i}, \ \ \ \ T^0_i =1, \ \ \ \ \
G_{ij}=1+u_{ij}\th_j\th_i  ,\label{GTNs}
\end{align}
where $\int$ ``integrates out'' all Grassmann numbers.

Similarly, $\Psi_f^*(\{m_i\})$ can be expressed as
\begin{align}
\Psi_f^*(\{m_i\})&=
\int \prod_i \bar T^{m_i}_i \prod_{ij}
(1+u^*_{ij}\bar \th_j\bar \th_i) ,
\nonumber\\
&=
\int \widetilde {\prod_i} \bar T^{m_i}_i \prod_{ij} \bar G_{ij} ,
\nonumber\\
\bar T^1_i &=\dd{\bar \th_i}, \ \ \ \ \bar T^0_i =1,
\nonumber\\
\bar G_{ij} &=1-u^*_{ij}\bar \th_j\bar \th_i  =1+u^*_{ij}\bar \th_i\bar \th_j  .
\end{align}
Note that $\prod_i$ and $\widetilde {\prod_i}$ have different orders:
\begin{align}
 \prod_i T^{m_i}_i \equiv T^{m_1}_1 T^{m_2}_2 T^{m_3}_3...\ \ \ \ \ \
 \widetilde {\prod_i}  T^{m_i}_i \equiv ...  T^{m_3}_3 T^{m_2}_2 T^{m_1}_1  .
\end{align}
Here we have used
the following identity
\begin{equation}
 \int [\prod_i  (\dd{\th_i})^{m_i}]
[\th_{i_1}\th_{i_2}...]
=  \int [\widetilde {\prod_i}  (\dd{\th_i})^{m_i} ]
[...\th_{i_2}\th_{i_1}]
\end{equation}

Thus the norm of the wave function is given by
\begin{align}
\label{normGTN}
 \<\Psi_f|\Psi_f\>&=
\sum_{\{m_i\}} \int
(\widetilde {\prod_i}
\bar T^{m_i}_i
\prod_{ij}
\bar G_{ij} )(
\prod_i T^{m_i}_i
\prod_{ij}
G_{ij} )
\nonumber\\
&=
\sum_{\{m_i\}} \int
\prod_i
\bar T^{m_i}_i
T^{m_i}_i
\prod_{ij}
\bar G_{ij}
G_{ij}
\nonumber\\
&=
\int
\prod_i
\v T_i
\prod_{ij}
\v G_{ij}
\end{align}
where
\begin{align}
 \v T_i=1+ \dd{\bar \th_i} \dd{\th_i} ,\
 \v G_{ij}= (1+u^*_{ij}\bar \th_i\bar \th_j) (1+u_{ij}\th_j\th_i) .
\end{align}

We may view $\v T_i$ as a dimension-1 tensor on the vertex $i$, and
$\v G_{ij}$ as a dimension-1 rank-2 tensor on the link $ij$.  Then
$\prod_i \v T^{m_i}_i \prod_{ij} \v G_{ij} $ can be viewed as the
tensor trace on such a tensor network.  Note that the tensors
contain Grassmann numbers and $\v T_i$'s always appear in front of
$\v G_{ij}$'s. We see that the norm of a fermion wave function can
be expressed as the tensor trace of a Grassmann tensor network.

\section{Grassmann tensor-network representation of generic
strongly correlated fermionic and bosonic states}

\begin{figure}
\begin{center}
\includegraphics[scale=0.7]{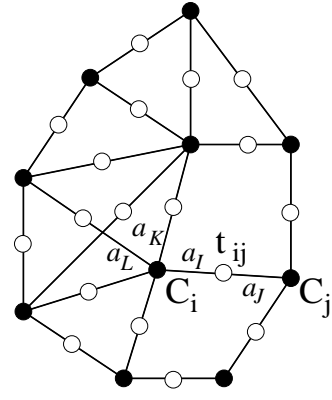}
\end{center}
\caption{ \label{grphCt} A strongly correlated system on a graph.
The physical degrees of freedom live on the vertices.  The vertices
(the filled dots) are labeled by $i$, $j$, \etc.  The open dots are
labeled by $ij$, \etc. The links connecting open and filled dots are
labeled by $I$, $J$, $K$, \etc. }
\end{figure}

The above result can be generalized to strongly correlated fermionic
states as well as strongly correlated bosonic states.  Let us assume
the physical degrees of freedom are localized on the vertices of a
graph (see Fig. \ref{rlatt}).  The vertices are labeled by $i$, $j$,
\etc.  The states on vertex $i$ are label by $m_i=1,2,...$ Let a
sign function $s(m_i)$ to have the following property.  $s(m_i)=1$
if the state $m_i$ is a bosonic state and $s(m_i)=-1$ if the state
$m_i$ is a fermionic state.  To construct a many-body wave function
$\Psi(\{m_i\})$, we introduce some fermion operators $\psi_i^{\al}$
on each vertex $i$ with $\al$ denotes the fermion species(such as
spin). We also use $ij$ \etc to label the open dots and $I$, $J$,
\etc to label the links between the open and filled dots in Fig.
\ref{grphCt}. Then we can construct $\Psi(\{m_i\})$ as
\begin{align}
\label{PsiTG}
\Psi(\{m_i\})=\sum_{\{a_I\}}
\<0| \prod_i C_{m_i;a_Ka_L...}
\prod_{\langle ij \rangle}
t_{ij;a_I a_J}
\ket{0} ,
\end{align}
where $K$, $L$, \etc in $C_{m_i;a_Ka_L...}$ label the links that
connect to the vertex $i$, and $I$, $J$ in $t_{ij;a_I a_J}$ label
the links that connect the vertex $i$, $j$ with the open dot $ij$.
All the link indices $a_K,a_L,a_I,a_J...=1,...,D$ are the inner
indices defined in the standard (bosonic) tesor product states(TPS).
Here $t_{ij;a_I b_J}$ only contains $\psi_i^{{\al\dag}}$ and
$\psi_j^{{\al\dag}}$ operators and $C_{m_i;a_Ka_L...}$ only contains
$\psi_i^{\al}$ operators. $t_{ij;a_Ia_J}$ always contains even
numbers of fermion operators. $C_{m_i;a_Ka_L...}$  contains an even
numbers of fermion operators if $s(m_i)=1$ and contains odd numbers
of fermion operators if $s(m_i)=-1$.

\begin{figure}
\begin{center}
\includegraphics[scale=0.7]{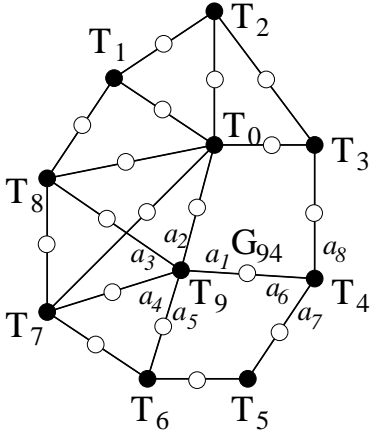}
\end{center}
\caption{ \label{tnTG} A tensor network formed by two kinds of
Grassmann tensors $T^{m_i}_{i;a_K a_L...}$ and $G_{ij;a_Ia_J}$ \etc.
The filled dots (the vertices) are labeled by $i$ which represent
$T^{m_i}_{i;a_K a_L...}$, and the open dots are labeled by $ij$
which represent $G_{ij;a_Ia_J}$.  The lines (labeled by $I$, $J$,
\etc) connecting the dots represent the indices $a_I$ of the
tensors.  For two tenors connected by a line, the values of the
associated indices  in the two tensor are set to be equal, and those
values are summed over in the tensor trace.  Such a summation is
represented by $\sum_{\{a_I\}} $ in \eq{PsiTG}. }
\end{figure}

Introducing the Grassmann numbers $\th_i^\al$ for each vertex, we
can express the many-body wave function as
\begin{align}
\Psi(\{m_i\})&= \sum_{\{a_I\}} \int \prod_i T^{m_i}_{i;a_Ka_L...}
\prod_{ij} G_{ij;a_I a_J} , \label{GTN}
\end{align}
where $T^{m_i}_{i;a_Ka_L...}$ is obtained from $C_{m_i;a_Ka_L...}$
by replacing $\psi_i^\al$ by $\dd{\th_i^\al}$ and $G_{ij;a_I a_J}$
is obtained from $t_{ij;a_I a_J}$ by replacing $\psi_i^{\al\dag}$ by
$\th_i^\al$.  So, $\Psi(\{m_i\})$ can be expressed as a tensor trace
over a Grassmann tensor network (see Fig. \ref{tnTG}).

Similarly
$\Psi^*(\{m_i\})$ can be expressed as
\begin{align}
\Psi^*(\{m_i\})&=
\sum_{\{a_I\}}
\int \widetilde {\prod_i} \bar T^{m_i}_{i; a_K a_L...}
\prod_{ij} \bar G_{ij; a_Ia_J} ,
\end{align}
where $\bar T^{m_i}_{i; a_K a_L...}$ is obtained from
$C^\dag_{m_i;a_Ka_L...}$ by replacing ${\psi_i^{\al}}^\dagger$ by
$\dd{\bar\th_i^\al}$ and $\bar G_{ij;a_Ia_J}$ is obtained from
$t^\dag_{ij;a_Ia_J}$ by replacing $\psi_i^{\al}$ by $\bar
\th_i^\al$. Essentially, \eq{GTN} is a fermionic generalization of
the standard (bosonic) tensor product states(TPS). If there is no
Grassmann number in $T,G$, it becomes the standard TPS (one can
further put $G_{ij;a_I a_J}=\delta_{a_I,a_J}$ if the tensor
contraction of inner indices are made over a trivial metric).

The norm of the wave function can be calculated in the same way:
\begin{align}
\label{normGTNg1}
& \<\Psi|\Psi\> =
\sum_{\{m_i\}, \{a_I \bar a_I\} }
\int
(\widetilde {\prod_i}
\bar T^{m_i}_{i;\bar a_K\bar a_L..}
\prod_{ij}
\bar G_{ij;\bar a_I\bar a_J} )\times
\nonumber\\
&\ \ \ \ \ \ \ \ \ \ \ \ \ \ \ \ \ \ \ \ \ \ \ \ \ \ \ \ \ \ \ \ \ \
(\prod_i
T^{m_i}_{i; a_Ka_L..}
\prod_{ij}
G_{ij;a_Ia_J} )
\nonumber\\
&=
\sum_{\{m_i\}, \{a_I \bar a_I\} }
\int
\prod_i
\bar T^{m_i}_{i;\bar a_K\bar a_L..}
T^{m_i}_{i; a_Ka_L..}
\prod_{ij}
\bar G_{ij;\bar a_I\bar a_J}
G_{ij;a_I a_J}
\nonumber\\
&=
\sum_{\{a_I \bar a_I\}}
\int
\prod_i
\v T_{i;a_K\bar a_K,a_L\bar a_L...}
\prod_{ij}
\v G_{ij;a_I\bar a_I,a_J\bar a_J}
\end{align}
where
\begin{align}
 \v T_{i;a_K\bar a_K,a_L\bar a_L...} &= \sum_m
\bar T^{m}_{i; \bar a_K\bar a_L...} T^{m}_{i; a_Ka_L...}
\nonumber\\
 \v G_{ij;a_I\bar a_I,a_J\bar a_J} &=
\bar G_{ij;\bar a_I\bar a_J} G_{ij;a_I a_J}.
\end{align}
Again, the norm is a tensor trace of a Grassmann tensor network (see
Fig. \ref{tnTG} where $T$, $G$ are replaced by $\v T$, $\v G$ and
each link is indexed by a pair $a_I \bar a_I$). We may combine the
pair of indices $(a_I\bar a_I)$ into one $p_I$, and rewrite the
above as:
\begin{align}
\label{normGTNg}
& \<\Psi|\Psi\> =
\sum_{\{p_I\}}
\int
\prod_i
\v T_{i;p_Kp_L...}
\prod_{ij}
\v G_{ij;p_Ip_J}
\end{align}

\section{Grassman tensor product states -- a ``bond'' form}
\label{bondform}

Although the Grassmann tensor network representations for the norm
of fermion wave function, \eq{normGTN} and \eq{normGTNg}, are simple
and compact, however, it is not easy to implement the
renormalization calculation for such a generic Grassmann tensor
network.  In this section, we will consider Grassmann tensor
networks that have a special form which makes the renormalization
calculation easier.  Although the Grassmann tensor networks have a
special form, they can still represent generic strongly correlated
fermionic states described by the more generic form \eq{normGTNg}.

\begin{figure}
\begin{center}
\includegraphics[scale=0.45]{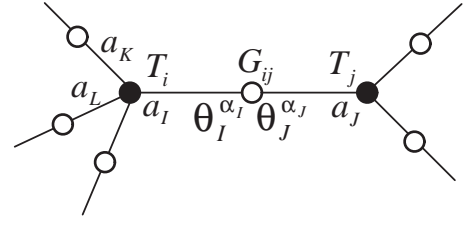}
\end{center}
\caption{ \label{TGij} the Grassmann number $\th^\al_{I}$ is
associated with the link $I$ that connects to vertex $i$. The tensor
$G_{ij}$ contains only Grassmann numbers $\th^\al_{I}$ and
$\th^\al_{J}$. }
\end{figure}

In the previous Grassmann tensor network, the Grassmann numbers on
vertex $i$, $\th^{\al_i}_i$, are labeled by $\al_i=1,2,...$ To
construct the more special form of the Grassmann tensor network, we
separate the Grassmann numbers on vertex $i$ into several groups,
one group for each link that connects the vertex $i$.  So each group
of Grassmann numbers is actually associated with a link labeled by
$I$. Thus it is more convenient to relabel the Grassmann numbers on
vertex $i$ as $\th^{\al_I}_{I}$, which correspond to the Grassmann
numbers on vertex $i$ and associated with the link $I$ that connects
to the vertex (see Fig.  \ref{TGij}).

With the new labeling scheme, we can specify the special form of the
Grassmann tensor network by requiring $G_{ij;a_I a_J}$ to only
contain the Grassmann numbers $\th^\al_I$ and $\th^\al_J$ (see Fig.
\ref{TGij}).  (Note that $G_{ij;a_Ia_J}$ still contain even numbers
of $\th^\al_I$ and $\th^\al_J$.) Under such scheme, the wavefunction
\eq{GTN} can be represented as:

\begin{align}
\Psi(\{m_i\})&= \sum_{\{a_I\}} \int \prod_i T^{m_i}_{i;a_Ka_L...}
\prod_{ij} G_{ij;a_I a_J} , \label{GTPS}
\end{align}
where
\begin{align}
&  T_{i;a_Ka_L...}^{m_i}= \sum_{\{l^{\al_K}_K\} \{l^{\al_L}_L\}..}
{\rm{T}}^{m_i;\{l^{\al_K}_K\}\{l^{\al_L}_L\}..}_{i;a_Ka_L..}\prod_{I\in
i} \widetilde{\prod_{\al_I}} (\dd \th^{\al_I}_{I})^{l^{\al_I}_I}
\nonumber\\
&  G_{ij;a_I a_J}=\sum_{\{l^{\al_I}_I\}\{l^{\al_J}_J\}}
{\rm{G}}^{\{l^{\al_I}_I\}\{l^{\al_J}_J\}}_{ij;a_I a_J} \prod_{\al_J}
(\th^{\al_J}_{J})^{l^{\al_J}_J} \prod_{\al_I}
(\th^{\al_I}_{I})^{l^{\al_I}_I}
\end{align}
Here $I \in i$ means the links that connect to vertex $i$. Note
that, in the expression of $G_{ij}$, we have assumed that the link
$I$ connects vertex $i$ with the open dot $ij$, and the link $J$
connects vertex $j$ with the open dot $ij$ (see Fig. \ref{TGij}). We
also note that $l^{\al_K}_K=$ 0 or 1 indicates the presence or the
absence of the Grassmann number $\dd \th^{\al_K}_K$.  So we may
interpret $l^{\al_K}_K$ as an ``occupation number of fermions''.Here
$\sum_{\{l^{\al_K}_K\}\{l^{\al_L}_L\}...} $ sums over all the
possible ``occupation'' distributions
$\{l^{\al_K}_K=0,1|\al_K=1,2,...\}$,
$\{l^{\al_L}_L=0,1|\al_L=1,2,...\}$,... Similarly as in \eq{GTN},
$\sum_{I\in i}\sum_{\al_I} l^{\al_I}_I =\text{ odd}$ represents a
fermionic state and $\sum_{I\in i}\sum_{\al_I} l^{\al_I}_I =\text{
even}$ represents a bosonic state. However, $\sum_{\al_I}
l^{\al_I}_I+ \sum_{\al_J} l^{\al_J}_J$ should always to be even.
Such type of representations for fermionic wave functions was first
introduced in \Ref{FrankfPEPS}.

In the new form, the norm of wave function can be expressed in the
same way as in \eq{normGTNg} with:

\begin{align}
 \v T_{i;a_K\bar a_K,a_L\bar a_L...} &= \sum_m
\bar T^{m}_{i; \bar a_K\bar a_L...} T^{m}_{i; a_Ka_L...}
\nonumber\\
 \v G_{ij;a_I\bar a_I,a_J\bar a_J} &=
\bar G_{ij;\bar a_I\bar a_J} G_{ij;a_I a_J}.\label{normTG}
\end{align}

Here $\bar T^{m}_{i; \bar a_K\bar a_L...}$ and $\bar G_{ij;\bar
a_I\bar a_J}$ are defined as:

\begin{align}
&  \bar T_{i;\bar a_K \bar a_L...}^{m_i}= \sum_{\{\bar l^{\al_K}_K\}
\{\bar l^{\al_L}_L\}..} \left[{\rm{T}}^{m_i;\{\bar
l^{\al_K}_K\}\{\bar l^{\al_L}_L\}..}_{i;\bar a_K \bar
a_L..}\right]^*\prod_{I\in i} {\prod_{\al_I}} (\dd \bar
\th^{\al_I}_{I})^{\bar l^{\al_I}_I}
\nonumber\\
&  \bar G_{ij;\bar a_I \bar a_J}=\sum_{\{\bar l^{\al_I}_I\}\{\bar
l^{\al_J}_J\}} \left[{\rm{G}}^{\{\bar l^{\al_I}_I\}\{\bar
l^{\al_J}_J\}}_{ij;\bar a_I \bar a_J}\right]^*
\widetilde{\prod_{\al_I}} (\bar \th^{\al_I}_{I})^{\bar
l^{\al_I}_I}\widetilde{\prod_{\al_J}} (\bar \th^{\al_J}_{J})^{\bar
l^{\al_J}_J}
\end{align}

By combining the pair of indices $(a_I\bar a_I)$ into one $p_I$ and
reordering those $d\th_I^{\al_I},d\bar
\th_I^{\al_I}$($\th_I^{\al_I},\bar \th_I^{\al_I}$), the tensors $\v
T$ and $\v G$ in \eq{normGTNg} can be further expanded as
\begin{align}
& \v T_{i;p_Kp_L...}= \sum_{\{n^{\al_K}_K\} \{n^{\al_L}_L\}..}
\cT^{\{n^{\al_K}_K\}\{n^{\al_L}_L\}..}_{i;p_Kp_L..} \prod_{I\in i}
\widetilde{\prod_{\al_I}} (\dd \eta^{\al_I}_{I})^{n^{\al_I}_I}
\nonumber\\
& \v G_{ij;p_I p_J}=\sum_{\{n^{\al_I}_I\}\{n^{\al_J}_J\}}
\cG^{\{n^{\al_I}_I\}\{n^{\al_J}_J\}}_{ij;p_I p_J} \prod_{\al_J}
(\eta^{\al_J}_{J})^{n^{\al_J}_J} \prod_{\al_I}
(\eta^{\al_I}_{I})^{n^{\al_I}_I}
\end{align}
where the group of Grassmann numbers $\{
\eta^{\al_I}_{I}|\al_I=1,2,... \}$ is the combination of $\{
\th^{\bt_I}_{I}|\bt_I=1,2,... \}$ and $\{
\bar\th^{\bt_I}_{I}|\bt_I=1,2,...  \}$.    Notice
$\cT^{\{n^{\al_K}_K\}\{n^{\al_L}_L\}...}_{i;p_Kp_L...}$ vanishes if
the total numbers of Grassmann numbers $\dd \eta^\al_I$ is odd and
$\cG^{\{n^{\al_I}_I\}\{n^{\al_J}_J\}}_{ij;p_I p_J} $ vanishes if the
total numbers of Grassmann numbers $\eta^\al_I,\eta^\al_J$ is odd:
\begin{align}
& \cT^{\{n^{\al_K}_K\}\{n^{\al_L}_L\}...}_{i;p_Kp_L...}
=0 ,\  \text{ if }
\sum_{I\in i}\sum_{\al_I} n^{\al_I}_I
=\text{ odd},
\\
& \cG^{\{n^{\al_I}_I\}\{n^{\al_J}_J\}}_{ij;p_I p_J}
=0 ,\  \text{ if }
\sum_{\al_I} n^{\al_I}_I+ \sum_{\al_J} n^{\al_J}_J
=\text{ odd}.
\nonumber
\end{align}

To gain some intuitive understanding of the new form of the
Grassmann tensor network, let us represent the free fermion wave
function $\Psi_f$ \eq{wvfree} using the new  form of Grassmann
tensor network:
\begin{align}
\label{PsifB}
& \Psi_f(\{m_i\})=\int \prod_i T^{m_i}_i \prod_{ij} G_{ij} ,
\nonumber\\
& T^1_i =\sum_{I \in i} \dd{\th_{I}}
, \ \ \  T^0_i =1, \ \ \ \
G_{ij} =1+u_{ij}\th_{J}\th_{I}  ,
\end{align}
where $\sum_{I \in i}$ sums over all the links that connect to
vertex $i$.  The norm of such a free fermion state is given by
\begin{align}
\label{normGTN1}
 \<\Psi_f|\Psi_f\>&=\int
\prod_i \v T_i \prod_{ij} \v G_{ij}
\nonumber\\
 \v T_i&=1+
[\sum_{K \in i} \dd\bar \th_{K}]
[\sum_{L \in i} \dd\th_L] ,
\nonumber\\
 \v G_{ij}&= (1+u^*_{ij}\bar \th_{I}\bar \th_{J})
(1+u_{ij}\th_{J}\th_{I}) .
\end{align}
Again, in the expression of $\v G_{ij}$, we have assumed that the link
$I$ connects the vertex $i$ with the open dot $ij$, and the link $J$
connects the vertex $j$ with the open dot $ij$.

We would like to stress that the above formulism is very general. It
can be used to express free fermion states, as well as strongly
correlated fermionic/bosonic states obtained from the projective
construction. It can even express strongly correlated
fermionic/bosonic states beyond the projective construction.  We
also would like to stress that the graphs discussed in this paper do
not have to be a real space lattice.

\section{Strongly correlated states from the projective construction}
\label{project}

As we have mentioned, the Grassmann tensor network can represent
very general strongly correlated fermionic/bosonic states. In this
section, we will concentrate on strongly correlated
fermionic/bosonic states obtained from a projective construction and
show that all of these states can be expressed as Grassmann tensor
product states.  The Grassmann tensor network for such projective
states takes a particularly simple form.

To construct the Grassmann tensor network for the projective states,
let us first construct the Grassmann tensor network for a ELF state:
\begin{align}
|\Psi_f \rangle=\exp\sb{\sum_{\langle ij \rangle}
(u_{ij})_{\al\bt}\psi_{j,\bt}^\dagger \psi_{i,\al}^\dagger } \ket{0}
\end{align}
where there can be several kinds of fermions labeled by $\al,\bt$ on
each vertex.  As discussed in section \ref{bondform}, the above ELF
state can be expressed in terms of Grassmann tensor product state:
\begin{align}
\label{PsifBg}
 \Psi_f(\{m_{i,\al}\}) &=
\<0| \prod_i \prod_\al (\psi_{i,\al})^{m_{i,\al}} |\Psi_f \rangle
\nonumber\\
& =\int \prod_i T^{\{m_{i,\al} \}}_{i} \prod_{ij} G_{ij} ,
\nonumber\\
 T^{\{m_{i,\al} \}}_{i} &=\prod_\al
(\sum_{I \in i} \dd{\th^\al_I})^{m_{i,\al}}
\nonumber\\
G_{ij} &= \exp\sb{\sum_{\langle ij \rangle}
(u_{ij})_{\al\bt}\th_{J}^\bt \th_{I}^\al }
\end{align}
where $m_{i,\al}=0,1$ describes the occupation of the
$\al^\text{th}$ fermion on vertex $i$.  $T^{\{m_{i,\al}
\}}_{i}$ is obtained from $\prod_\al
(\psi_{i,\al})^{m_{i,\al}} $ by replacing $\psi_{i,\al}$ by
$\sum_{I \in i} \dd{\th^\al_I}$ where $\sum_{I \in i}$ sums
over all the links that connects to the vertex $i$.
$G_{ij}$ is obtained from $\exp\sb{\sum_{\langle ij \rangle}
(u_{ij})_{\al\bt}\psi_{j,\bt}^\dagger \psi_{i,\al}^\dagger }
$ by replacing $\psi_{j,\bt}^\dag\psi_{i,\al}^\dag$ by
$\th_{J}^\bt \th_{I}^\al$ where $I$ and $J$ label the links
that connects to the open dot $ij$ between the two vertices
$i$ and $j$.  Note that \eq{PsifBg} is a generalization of
\eq{PsifB}.

To obtain a strongly correlated fermionic state from the above ELF
state, for example, we may assume $\al=1,2,3$ and consider the
following many-body wave function:
\begin{align}
 \Psi_\text{corr}(\{m_i\}) &=
\<0| \prod_i (c_i)^{m_i} |\Psi_f \rangle , & c_i &= \prod_{\al=1}^3
\psi_{i,\al}  ,
\end{align}
where $m_i=0,1$.  We note that $\Psi_\text{corr}(\{m_i\})$ is a
many-body wave function for a spinless fermion system.  It is a
strongly correlated projective state which is very different from
any ELF state.  Such a state can be expressed in terms of Grassmann
tensor network
\begin{align}
\label{tncorr} & \Psi_\text{corr}(\{m_i\}) = \int \prod_i
T^{m_i}_{i} \prod_{ij} G_{ij} ,
\\
& T^{m_i}_{i} =\Big(\prod_{\al=1}^3 \sum_{I \in i}
\dd{\th^\al_I}\Big)^{m_i} ,\ \ G_{ij} = \exp\sb{\sum_{\langle ij
\rangle} (u_{ij})_{\al\bt}\th_{J}^\bt \th_{I}^\al } \nonumber
\end{align}

Similarly, to obtain a strongly correlated hardcore bosonic state
(such as spin liquid state) using projective construction, for
example, we may assume $\al=1,2$ and consider the following
many-body wave function:
\begin{align}
 \Psi_\text{spin}(\{m_i\}) &=
\<0| \prod_i (b_i)^{m_i} |\Psi_f \rangle , & b_i &= \psi_{i,1}
\psi_{i,2} ,
\end{align}
where $m_i=0,1$.  $\Psi_\text{spin}(\{m_i\})$ is a many-body wave
function for a hardcore boson system.  It can also be viewed as a
wave function for a spin-1/2 system.  Such a state can be expressed
in terms of Grassmann tensor network
\begin{align}
\label{tnspin} & \Psi_\text{spin}(\{m_i\}) = \int \prod_i
T^{m_i}_{i} \prod_{ij} G_{ij} ,
\\
& T^{m_i}_{i} =\Big(\prod_{\al=1}^2 \sum_{I \in i}
\dd{\th^\al_I}\Big)^{m_i} ,\ \ G_{ij} = \exp\sb{\sum_{\langle ij
\rangle} (u_{ij})_{\al\bt}\th_{J}^\bt \th_{I}^\al } . \nonumber
\end{align}

Both strongly correlated states $\Psi_\text{corr}(\{m_i\})$ and
$\Psi_\text{spin}(\{m_i\})$ are parameterized by
$(u_{ij})_{\al\bt}$. We may view $(u_{ij})_{\al\bt}$ as variational
parameters.  After finding $(\bar u_{ij})_{\al\bt}$ that minimize
the average energy, we obtain the approximated ground state.  We can
also obtain the low energy effective theory from the form of the
ansatz $(\bar u_{ij})_{\al\bt}$.

Finally we would like to point out that the ways to construct ELF
states and projective states discussed in sections \ref{ELF} and
\ref{project} are simple but not efficient. The Grassmann tensor
product states derived in that way will usually contain long range
connections, which is not necessary in special
cases\cite{FrankfPEPS}.

\section{Fermion coherent state representation}
In above sections, we have represented the Grassmann tensor network
wavefunctions under the Fock basis. Although the Fock basis
representation is simple and straightforward to derive, however,
because of the anticommutating relations for different Grassmann
numbers, the wavefunctions depend on the ordering the local
Grassmann tensors $ T^{m_i}_{i;a_Ka_L...}$ and are inconvenient for
simulating physical quantities for fermion systems.

In this section, we would like to introduce the fermion coherent
state representation for Grassmann tensor product states. We show
the fermion wavefunctions in this basis are independent of the
ordering of local Grassmann tensors. To see this explicitly, let us
consider a simple spinless fermion tensor product state:
\begin{align}
|\Psi \> = \sum_{\{m_i\}} \sum_{\{a_I\}} \int \prod_i
[c_i^\dagger]^{m_i} T^{m_i}_{i;a_Ka_L...} \prod_{ij} G_{ij;a_I
a_J}|0\> , \label{SGTPS}
\end{align}
where $m_i=0,1$ represents the fermion occupation numbers. It is
easy to check that we can derive the Grassmann tensor product
wavefunction \eq{GTPS} under the following Fock basis:

\begin{align}
 \prod_i (c_i)^{m_i} |0\rangle\equiv  (c_1^\dagger)^{m_1} (c_2^\dagger)^{m_2}
(c_3^\dagger)^{m_3} ...|0\rangle
\end{align}

The over complete fermion coherent state basis is defined as:
\begin{align}
|\eta\rangle \equiv \prod_i (1-\eta_i c_i^{\dagger}) |0\rangle,
\end{align}
with closure relation:
\begin{align}
\int\prod_i \dd \eta_i^* \dd \eta_i (1-\eta_i^*\eta_i)|\eta \rangle
\langle \eta |=1.\label{closure}
\end{align}
It is easy to derive the wavefunction for Grassmann tensor product
state \eq{SGTPS} under such a basis:

\begin{align}
\langle \eta|\Psi \> = \sum_{\{m_i\}} \sum_{\{a_I\}} \int \prod_i
{\eta_i^*}^{m_i} T^{m_i}_{i;a_Ka_L...} \prod_{ij} G_{ij;a_I a_J} ,
\label{CGTPS}
\end{align}

If we redefine the Grassmann tensor $ T^{m_i}_{i;a_Ka_L...}$ as:
\begin{align}
 T_{i;a_Ka_L...}^{m_i}= \sum_{\{l^{\al_K}_K\} \{l^{\al_L}_L\}..}
{\rm{T}}^{m_i;\{l^{\al_K}_K\}\{l^{\al_L}_L\}..}_{i;a_Ka_L..}{\eta_i^*}^{m_i}\prod_{I\in
i} \widetilde{\prod_{\al_I}} (\dd \th^{\al_I}_{I})^{l^{\al_I}_I},
\label{Tnew}
\end{align}
then the wavefunction can be represented as:

\begin{align}
\Psi_{\rm{coh}}(\eta^*)=\sum_{\{m_i\},\{a_I\}} \int \prod_i
T^{m_i}_{i;a_Ka_L...} \prod_{ij} G_{ij;a_I a_J} , \label{NCGTPS}
\end{align}
Notice under the new definition, $ T_{i;a_Ka_L...}^{m_i}$ always
contain \emph{even} number of Grassmann numbers hence the definition
of the wavefunction \eq{NCGTPS} is independent of how we order those
local Grassmann tensors $ T_{i;a_Ka_L...}^{m_i}$.

It is very convenient to use the above wavefunction to calculate
local physical quantities. The only thing we need to take care is
the over complete nature of the basis, hence a proper measure is
needed when we calculate the inner product for a wavefunction. For
example, the norm of the wavefunction \eq{NCGTPS} can be calculated
as:

\begin{align}
&\int\prod_i \dd \eta_i^* \dd \eta_i
(1-\eta_i^*\eta_i)\Psi_{\rm{coh}}^*(\eta)\Psi_{\rm{coh}}(\eta^*)\nonumber\\&=
\sum_{\{p_I\}} \int \prod_i \v T_{i;p_Kp_L...} \prod_{ij} \v
G_{ij;p_Ip_J},
\end{align}
which is exactly the same result as we derived before(tensors $\v T$
and $\v G$ are defined in Eq. (\ref{normTG})).

Other physical quantities like the average energy can also be easily
calculated in a similar way. For any local operators containing
$c^\dagger$ and $c$, we only need to replace $c^\dagger$ with
$\eta^*$ and $c$ with $\eta$, then integrate out the Grassmann
number respect to a proper measure as we do for calculating the
norm. For example, the nearest neighbor paring term $c_i^\dagger
c_j^\dagger$ can be expressed as:

\begin{widetext}
\begin{align}
\label{pairGTPS} \<\Psi| c_i^\dagger c_j^\dagger |\Psi\> &=\int\dd
\eta_i^* \dd \eta_i (1-\eta_i^*\eta_i)\dd \eta_j^* \dd \eta_j
(1-\eta_j^*\eta_j)\prod_{i^\prime} \dd \eta_{i^\prime}^* \dd
\eta_{i^\prime}
(1-\eta_{i^\prime}^*\eta_{i^\prime})\eta_i^*\eta_j^*\Psi_{\rm{coh}}^*(\eta)\Psi_{\rm{coh}}(\eta^*)
\nonumber\\ &= -\sum_{ a_M \bar a_M..a_K \bar a_K.. }
\sum_{\{m_{i^\prime}\}, \{a_{I^\prime} \bar a_{I^\prime}\} } \int
\bar T^1_{j;\bar a_M \bar a_N..} \bar T^1_{i;\bar a_K \bar
a_L..}\bar G_{ij;\bar a_{I}\bar a_{J}} T^0_{i; a_K a_L..}T^0_{j; a_M
a_N..} G_{ij;a_{I} a_{J}}  \nonumber\\ &\times \widetilde
{\prod_{i^\prime}} \bar T^{m_{i^\prime}}_{{i^\prime};\bar
a_{K^\prime}\bar a_{L^\prime}..} \prod_{{i^\prime}{j^\prime}} \bar
G_{{i^\prime}{j^\prime};\bar a_{I^\prime}\bar a_{J^\prime}}
\prod_{i^\prime} T^{m_{i^\prime}}_{{i^\prime};
a_{K^\prime}a_{L^\prime}..} \prod_{{i^\prime}{j^\prime}}
G_{{i^\prime}{j^\prime};a_{I^\prime}a_{J^\prime}}
\nonumber\\
&= \sum_{\{a_{I^\prime} \bar a_{I^\prime}\}} \int \v T^\prime
_{i;a_K \bar a_K,a_{L}\bar a_{L}...} \v T^{\prime\prime} _{j;a_M
\bar a_M,a_{N}\bar a_{N}...} \v G_{i j ;a_{I}\bar a_{I},a_{J}\bar
a_{J}} \prod_{i^\prime} \v T_{i^\prime;a_{K^\prime}\bar
a_{K^\prime},a_{L^\prime}\bar a_{L^\prime}...} \prod_{i^\prime
j^\prime } \v G_{i^\prime j^\prime ;a_{I^\prime}\bar
a_{I^\prime},a_{J^\prime}\bar a_{J^\prime}},
\end{align}
\end{widetext}
where the impurity tensors $\v T^\prime$ and $\v T^{\prime\prime}$
are defined as:
\begin{align}
 \v T_{i;a_K\bar a_K,a_L\bar a_L...}^\prime &=
\bar T^{1}_{i; \bar a_K\bar a_L...} T^{0}_{i; a_Ka_L...}
\nonumber\\
 \v T_{j;a_M\bar a_M,a_N\bar a_N...}^{\prime\prime} &=
\bar T^{1}_{j; \bar a_M\bar a_N...} T^{0}_{j;
a_Ma_N...}.\label{impurityT}
\end{align}
and the indices $i^\prime$ denote other sites beside $i$ and $j$.

In the last line we omit the minus sign because the two impurity
tensors $\v T^\prime$ and $\v T^{\prime\prime}$ contain \emph{odd}
number of Grassmann numbers hence they \emph{anticommute} with each
other.

In conclusion, the norm of a Grassmann tensor product state can be
expressed as a tensor trace of uniform Grasmann tensor-net and other
local physical quantities such as energy can be expressed as a
tensor trace of Grasmann tensor-nets with impurity tensors. Here we
have already seen that the fermion coherent state representation is
extremely convenient for expressing the norm and physical quantities
of Grassmann tensor product states as tensor traces of Grassmann
tensor-nets. Essentially, such representations are the most natural
representations for Grassmann tensor product state and provide us a
deep insight into what is a fermion wavefunction. Actually, a
fermion wavefunctions should be described Grassmann numbers rather
than complex numbers, a detail discussions will be represented in
our future publications.

\section{The coarse graining transformation of Grassmann tensor
network}

Expressing the norm and physical quantities of a strongly correlated
wave function in terms of a tensor trace over a Grassmann tensor
network is a very formal exercise.  Calculating such a tensor trace
directly is a exponential hard problem.  So it seems that we gain
nothing from writing the norm in the form of tensor trace.

However, if we only want to evaluate the norm and physical
quantities approximately, then there is a polynomial way to do so in
terms of the Grassmann tensor network. The basic idea is to perform
a coarse graining transformation of the tensor network which can
simplify the tensor network into one with only a few
tensors.\cite{LN0701,GLWtergV}

In this section, we will explain how to apply a coarse graining
transformation to a Grassmann tensor network.  We will discuss three
basic moves on a honeycomb lattice, using rank-two, rank-three and
rank-four tensors as examples.

\begin{figure}
\begin{center}
\includegraphics[scale=0.7]{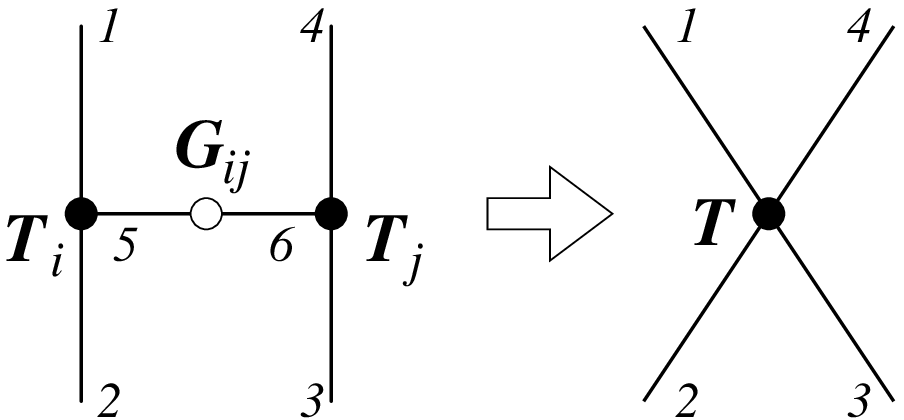}
\end{center}
\caption{ \label{TiTjT} We note that the Grassmann numbers
$\th^{\al_I}_{I}$ and $\dd \th^{\al_I}_{I}$ are associated with the
links the connect the $\v G$ tensors and $\v T$ tensors.  Here we
label the links with $I=$ 1, 2, 3, 4, 5, and 6.  The tensor $\v
T_{i;p_5p_1p_2}$ contains Grassmann numbers $\{ \dd
\eta^{\al_5}_{5}, \dd \eta^{\al_1}_{1}, \dd \eta^{\al_2}_{2} \} $.
The tensor $\v T_{j;p_6p_3p_4}$ contains Grassmann numbers $\{
\dd\eta^{\al_6}_{6}, \dd\eta^{\al_3}_{3}, \dd\eta^{\al_4}_{4}\} $.
The tensor $\v T$ contains Grassmann numbers $\{ \dd \eta^{\al_1}_1,
\dd \eta^{\al_2}_2, \dd \eta^{\al_3}_3, \dd \eta^{\al_4}_4\} $.  The
tensor $\v G_{ij}$ contains Grassmann numbers $\{ \eta^{\al_5}_5,
\eta^{\al_6}_6 \} $. }
\end{figure}

In the first move, we combine two rank-three tensors $\v T_i,\ \v T_j$ and one
rank-two tensor $\v G_{ij}$ into one rank-four tensor $\v T$ (see Fig.
\ref{TiTjT}).
We have
\begin{equation}
\label{TTTG}
 \v T_{p_1p_2p_3p_4}=\sum_{p_5p_6} \int_{ij}
\v T_{i;p_5p_1p_2}
\v T_{j;p_6p_3p_4}
\v G_{ij;p_5p_6}
\end{equation}
where $\int_{ij}$ only integrate over $\eta^{\al_5}_i$ and
$\eta^{\al_6}_j$.  If we expand $\v T_i$, $\v T_j$, $\v G_{ij}$, and
$\v T_i$, we get
\begin{widetext}
\begin{align}
\label{expTG}
& \v T_{i;p_5p_1p_2}=
\sum_{\{n^{\al_5}_5\}\{n^{\al_1}_1\} \{n^{\al_2}_2\}}
\cT^{\{n^{\al_5}_5\}\{n^{\al_1}_1\}\{n^{\al_2}_2\}}_{i;p_5p_1p_2}
\widetilde{\prod_{\al_5}} (\dd \eta^{\al_5}_{5})^{n^{\al_5}_5}
\widetilde{\prod_{\al_1}} (\dd \eta^{\al_1}_{1})^{n^{\al_1}_1}
\widetilde{\prod_{\al_2}} (\dd \eta^{\al_2}_{2})^{n^{\al_2}_2}
\nonumber\\
& \v T_{j;p_6p_3p_4}=
\sum_{\{n^{\al_6}_6\}\{n^{\al_3}_3\} \{n^{\al_4}_4\}}
\cT^{\{n^{\al_6}_6\}\{n^{\al_3}_3\}\{n^{\al_4}_4\}}_{j;p_6p_3p_4}
\widetilde{\prod_{\al_6}} (\dd \eta^{\al_6}_{6})^{n^{\al_6}_6}
\widetilde{\prod_{\al_3}} (\dd \eta^{\al_3}_{3})^{n^{\al_3}_3}
\widetilde{\prod_{\al_4}} (\dd \eta^{\al_4}_{4})^{n^{\al_4}_4}
\nonumber\\
& \v T_{p_1p_2p_3p_4}=
\sum_{\{n^{\al_1}_1\}\{n^{\al_2}_2\}\{n^{\al_3}_3\} \{n^{\al_4}_4\}}
\cT^{\{n^{\al_2}_2\}\{n^{\al_2}_2\}\{n^{\al_3}_3\}\{n^{\al_4}_4\}}
_{p_1p_2p_3p_4}
\widetilde{\prod_{\al_1}} (\dd \eta^{\al_1}_{1})^{n^{\al_1}_1}
\widetilde{\prod_{\al_2}} (\dd \eta^{\al_2}_{2})^{n^{\al_2}_2}
\widetilde{\prod_{\al_3}} (\dd \eta^{\al_3}_{3})^{n^{\al_3}_3}
\widetilde{\prod_{\al_4}} (\dd \eta^{\al_4}_{4})^{n^{\al_4}_4}
\nonumber\\
& \v G_{ij;p_5 p_6}=\sum_{\{n^{\al_5}_5\}\{n^{\al_6}_6\}}
\cG^{\{n^{\al_5}_5\}\{n^{\al_6}_6\}}_{ij;p_5 p_6}
\prod_{\al_5} (\eta^{\al_5}_{5})^{n^{\al_5}_5}
\prod_{\al_6} (\eta^{\al_6}_{6})^{n^{\al_6}_6}
\end{align}
We find that
\begin{align}
\cT^{\{n^{\al_2}_2\}\{n^{\al_2}_2\}\{n^{\al_3}_3\}\{n^{\al_4}_4\}}
_{p_1p_2p_3p_4}
=
\sum_{p_5p_6}
\sum_{\{n^{\al_5}_5\}\{n^{\al_6}_6\}}
\cT^{\{n^{\al_5}_5\}\{n^{\al_1}_1\}\{n^{\al_2}_2\}}_{i;p_5p_1p_2}
\cG^{\{n^{\al_5}_5\}\{n^{\al_6}_6\}}_{ij;p_5 p_6}
\cT^{\{n^{\al_6}_6\}\{n^{\al_3}_3\}\{n^{\al_4}_4\}}_{j;p_6p_3p_4}
\end{align}
\end{widetext}
This allows us to calculate $\v T$ from $\v T_i$, $\v T_j$, and $\v
G_{ij}$.

\begin{figure}
\begin{center}
\includegraphics[scale=0.7]{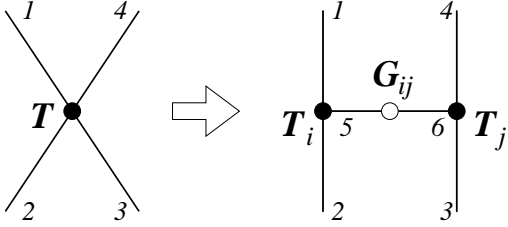}
\end{center}
\caption{
\label{TTiTj}
Split one rank-four tensor into two rank-three tensors and a rank-two
tensor.
}
\end{figure}

The second basic move splits the rank-four tensor $\v T$ into two
rank-three tensors $\v T_i$, $\v T_j$ and a rank-two tensor $\v
G_{ij}$ (see Fig.  \ref{TTiTj}).  We first rewrite
$\cT^{\{n^{\al_2}_2\}\{n^{\al_2}_2\}\{n^{\al_3}_3\}\{n^{\al_4}_4\}}
_{p_1p_2p_3p_4} $ as (say, using the
singular-value-decomposition(SVD) method discussed in \Ref{LN0701})
\begin{align}
 \cT^{\{n^{\al_2}_2\}\{n^{\al_2}_2\}\{n^{\al_3}_3\}\{n^{\al_4}_4\}}
_{p_1p_2p_3p_4}
=
\sum_{q}
\cS^{\{n^{\al_1}_1\}\{n^{\al_2}_2\}}_{i;qp_1p_2}
\cS^{\{n^{\al_3}_3\}\{n^{\al_4}_4\}}_{j;qp_3p_4} .
\end{align}
The above decomposition can be rewritten as
\begin{align}
&\ \ \ \cT^{\{n^{\al_2}_2\}\{n^{\al_2}_2\}\{n^{\al_3}_3\}\{n^{\al_4}_4\}}
_{p_1p_2p_3p_4}
\nonumber\\
&=
\sum_{p_5p_6}
\cT^{\{n_5\}\{n^{\al_1}_1\}\{n^{\al_2}_2\}}_{i;p_5p_1p_2}
\cG^{\{n_5\}\{n_6\}}_{ij;p_5 p_6}
\cT^{\{n_6\}\{n^{\al_3}_3\}\{n^{\al_4}_4\}}_{j;p_6p_3p_4}
\end{align}
where
\begin{align}
\label{GTSTS}
\cG^{\{n_5\}\{n_6\}}_{ij;p_5 p_6}
&= \del_{p_5p_6} \del_{n_5n_6},
\nonumber\\
\cT^{\{n_5\}\{n^{\al_1}_1\}\{n^{\al_2}_2\}}_{i;qp_1p_2}
&=
\cS^{\{n^{\al_1}_1\}\{n^{\al_2}_2\}}_{i;qp_1p_2} ,
\nonumber\\
\cT^{\{n_6\}\{n^{\al_3}_3\}\{n^{\al_4}_4\}}_{j;qp_3p_4}
&=
\cS^{\{n^{\al_3}_3\}\{n^{\al_4}_4\}}_{j;qp_3p_4} ,
\end{align}
and
\begin{align}
 n_5 &= \sum_{\al_1} n^{\al_1}_1 +\sum_{\al_2} n^{\al_2}_2 \text{ mod } 2
\nonumber\\
 n_6 &= \sum_{\al_3} n^{\al_3}_3 +\sum_{\al_4} n^{\al_4}_4 \text{ mod } 2.
\end{align}
We see that $n_5$ and $n_6$ are completely fixed by $ n^{\al_I}_I$,
$I=1,2,3,4$.  \Eqn{GTSTS} define three Grassmann tensors (see
\eq{expTG})
\begin{widetext}
\begin{align}
& \v T_{i;p_5p_1p_2}=
\sum_{\{n_5\}\{n^{\al_1}_1\} \{n^{\al_2}_2\}}
\cT^{\{n_5\}\{n^{\al_1}_1\}\{n^{\al_2}_2\}}_{i;p_5p_1p_2}
(\dd \eta_{5})^{n_5}
\widetilde{\prod_{\al_1}} (\dd \eta^{\al_1}_{1})^{n^{\al_1}_1}
\widetilde{\prod_{\al_2}} (\dd \eta^{\al_2}_{2})^{n^{\al_2}_2}
\nonumber\\
& \v T_{j;p_6p_3p_4}=
\sum_{\{n_6\}\{n^{\al_3}_3\} \{n^{\al_4}_4\}}
\cT^{\{n_6\}\{n^{\al_3}_3\}\{n^{\al_4}_4\}}_{j;p_6p_3p_4}
(\dd \eta_{6})^{n_6}
\widetilde{\prod_{\al_3}} (\dd \eta^{\al_3}_{3})^{n^{\al_3}_3}
\widetilde{\prod_{\al_4}} (\dd \eta^{\al_4}_{4})^{n^{\al_4}_4}
\nonumber\\
& \v G_{ij;p_5 p_6}=(1+ \eta_{5} \eta_{6})\del_{p_5p_6}
\end{align}
\end{widetext}
This way, we split the tensor $\v T$ into to the above three
tensors, since $\v T$ can be expressed in terms of the three
tensors as in \eq{TTTG}.  It is interesting to note that $\v
G_{ij}$ reduces to a very simple form after one step of
second move.

\begin{figure}
\begin{center}
\includegraphics[scale=0.7]{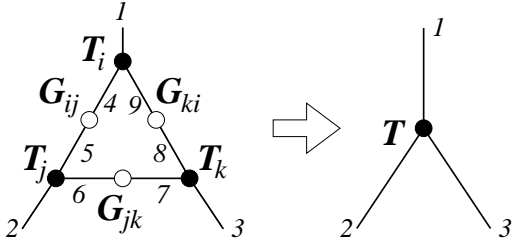}
\end{center}
\caption{
\label{TijkT}
Combine three rank-three tensors
and three rank-two tensors into one rank-three tensor.
}
\end{figure}

In the third move, we combine three rank-three tensors $\v T_i,\ \v
T_j,\ \v T_k$ and three rank-two tensors $\v G_{ij},\ \v G_{jk},\ \v
G_{ki}$ into one rank-three tensor $\v T$ (see Fig.  \ref{TijkT}).
If we expand the tensors $ \v T_{i;p_1p_4p_9},\ \v T_{j;p_2p_6p_5},\
\v T_{k;p_3p_8p_7}$ and  $\v G_{ij;p_4p_5},\ \v G_{jk;p_6p_7},\ \v
G_{ki;p_8p_9}$, we obtain the following coefficients
$\cT^{\{n^{\al_1}_1\}\{n^{\al_4}_4\}\{n^{\al_9}_9\}}_{i;p_1p_4p_9}$,
$\cT^{\{n^{\al_2}_2\}\{n^{\al_6}_6\}\{n^{\al_5}_5\}}_{j;p_2p_6p_5}$,
$\cT^{\{n^{\al_3}_3\}\{n^{\al_8}_8\}\{n^{\al_7}_7\}}_{k;p_3p_8p_7}$,
$\cG^{\{n^{\al_4}_4\}\{n^{\al_5}_5\}}_{ij;p_4p_5}$,
$\cG^{\{n^{\al_6}_6\}\{n^{\al_7}_7\}}_{jk;p_6p_7}$,
$\cG^{\{n^{\al_8}_8\}\{n^{\al_9}_9\}}_{ki;p_8p_9}$. The coefficients
of the resulting tensor $\v T_{p_1p_2p_3}$ are given by
\begin{widetext}
\begin{align}
\cT^{\{n^{\al_1}_1\}\{n^{\al_2}_2\}\{n^{\al_3}_3\}}_{p_1p_2p_3} &=
\sum_{p_4p_5p_6p_7p_8p_9} \sum_{ \{n^{\al_4}_4\} } \sum_{
\{n^{\al_5}_5\} } \sum_{ \{n^{\al_6}_6\} } \sum_{ \{n^{\al_7}_7\} }
\sum_{ \{n^{\al_8}_8\} } \sum_{ \{n^{\al_9}_9\} } (-)^{
(\sum_{\al_8} n^{\al_8}_8) (\sum_{\al_9} n^{\al_9}_9) } \times
\nonumber\\
& \ \ \ \ \
\cT^{\{n^{\al_1}_1\}\{n^{\al_4}_4\}\{n^{\al_9}_9\}}_{i;p_1p_4p_9}
\cT^{\{n^{\al_2}_2\}\{n^{\al_6}_6\}\{n^{\al_5}_5\}}_{j;p_2p_6p_5}
\cT^{\{n^{\al_3}_3\}\{n^{\al_8}_8\}\{n^{\al_7}_7\}}_{k;p_3p_8p_7}
\cG^{\{n^{\al_4}_4\}\{n^{\al_5}_5\}}_{ij;p_4p_5}
\cG^{\{n^{\al_6}_6\}\{n^{\al_7}_7\}}_{jk;p_6p_7}
\cG^{\{n^{\al_8}_8\}\{n^{\al_9}_9\}}_{ki;p_8p_9}
\end{align}
\end{widetext}

Just as in the bosonic TERG case\cite{GLWtergV}, the above coarse
graining transformation can also be generalized onto other plainer
graphs, such as square, kagome, triangular lattice, etc. However, on
generic graphs, especially those graphs with long range connections,
the tensor contraction still can be exponentially hard in most
cases.

\section{Example: Free fermion state on a honeycomb lattice}

\begin{figure}
\centerline{
\includegraphics[width=1.6in, height=1.3in]{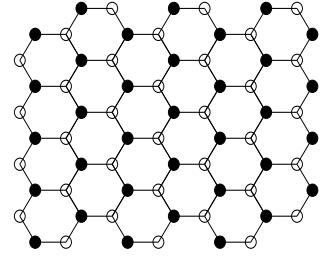}
}
\caption{
\label{hclatt}
A honeycomb lattice and its two sublattices.
The distance between two nearest neighboring sites
is chosen to be 1.
}
\end{figure}

To test the coarse graining procedure for the Grassmann tensor
network discussed above, let us study a simple example of a free
fermion state and its tensor network representations.  We assume
fermions live on the vertices of a honeycomb lattice.  Let us
consider the pairing state
\begin{equation}
\label{PsifHC}
 |\Psi_f\>=\e^{P}|0\>
\end{equation}
on a honeycomb lattice.  Here the pairing operator is given by
\begin{equation*}
 P=
\sum_{\<\v i\v j\>}   u c^\dag_{\v i} c^\dag_{\v j}
=\sum_{\v i \in A, a=1,2,3}   u c^\dag_{\v i} c^\dag_{\v i+{\v\del}_a}
\end{equation*}
where $\v i$ and $\v j$ label sites and $\<\v i\v j\>$ labels the
nearest-neighbor links of the honeycomb lattice. We know that the
sites of the  honeycomb lattice can be divided into two sub-lattice:
A and B (see Fig. \ref{hclatt}). The pairing of the fermions is only
between the two different sublattices.  Here we have used the
convention that in the link label $\<\v i\v j\>$, $\v i$ be long to
the A-sublattice and $\v j$ belong to the B-sublattice.  The three
vectors $\v\del_a$, $a=1,2,3$, are the three vectors that connect a
A-site to to its three nearest neighboring B-sites.

Introduce
\begin{align*}
 c_A(\v k)&=\sqrt\frac{2}{N}\sum_{\v i \in A} \e^{-\imth \v k\cdot \v i} c_{\v i}
\nonumber\\
 c_B(\v k)&=\sqrt\frac{2}{N}\sum_{\v i \in B} \e^{-\imth \v k\cdot \v i} c_{\v i}
 =\sqrt\frac{2}{N}\sum_{\v i \in A} \e^{-\imth \v k\cdot (\v i+\v\del_a)} c_{\v i+\v\del_a}
\end{align*}
where $N$ is the total number of lattice sites,
$\sum_{\v i \in A}$ sums over the sites in the A-sublattice and
$\sum_{\v i \in B}$ over the B-sublattice.
We can rewrite $P$ as
\begin{align*}
 P&=\sum_{\v k} c^\dag_A(-\v k) \al_{\v k} c^\dag_B(\v k)
\end{align*}
with
\begin{align*}
 \al_{\v k}&=
 u[\e^{-\imth k_x}
+  \e^{-\imth (-\frac{\sqrt 3 k_y}{2} -\frac{k_x}2 )}
+  \e^{-\imth (\frac{\sqrt 3 k_y}{2} -\frac{k_x}2 )}]
.
\end{align*}
Therefore
\begin{align}
 |\Psi_f\>&=
\prod_{\v k}
[
1+
\al_{\v k}
c^\dag_A(-\v k)
c^\dag_B(\v k)
] |0\>
.
\end{align}
Let us rewrite
\begin{align}
&\ \ \ [
1+
\al_{\v k}
c^\dag_A(-\v k)
c^\dag_B(\v k)
] |0\>
\nonumber\\
&\propto
[
  v_{\v k} c^\dag_A(-\v k)
+ u_{\v k} c_B(\v k)
]
[
  u_{\v k} c_A(-\v k)
- v_{\v k} c^\dag_B(\v k)
]
|0\>
\end{align}
where
\begin{align}
|u_{\v k}|^2+|v_{\v k}|^2=1
,\ \ \ \
\al_{\v k}=v_{\v k}/u_{\v k}
.
\end{align}
We find
\begin{align}
 v_{\v k}=\frac{\al_{\v k}}{\sqrt{1+|\al_{\v k}|^2}} , \ \ \ \
 u_{\v k}= \frac{1}{\sqrt{1+|\al_{\v k}|^2}} .
\end{align}
Let
\begin{align}
\psi_{1,-\v k}&=
[
  u_{\v k} c_A(-\v k)
- v_{\v k} c^\dag_B(\v k)
]
,
\nonumber\\
\psi_{2,\v k}&=
[
  v_{\v k} c^\dag_A(-\v k)
+ u_{\v k} c_B(\v k)
],
\end{align}
which satisfy the standard commutation relation for fermion operators.
By construction, we see that
\begin{align}
 \psi_{1,\v k}|\Phi_f\>
= \psi_{2,\v k}|\Phi_f\>=0 .
\end{align}
Thus $|\Psi_f\>$ is the ground state of the following quadratic
Hamiltonian
\begin{align}
 H &= \sum_{\v k}
(1+|\al_{\v k}|^2)
( \psi^\dag_{1,\v k} \psi_{1,\v k} +\psi^\dag_{2,\v k} \psi_{2,\v k})
\nonumber\\
&=
\sum_{\v k}
[ c^\dag_A(-\v k) - \al^*_{\v k} c_B(\v k) ]
[ c_A(-\v k) - \al_{\v k} c^\dag_B(\v k) ]
\nonumber\\
&\ \ +\sum_{\v k}
[ \al^*_{\v k} c_A(-\v k)  +  c^\dag_B(\v k) ]
[ \al_{\v k} c^\dag_A(-\v k)  +  c_B(\v k) ]
\nonumber\\
&=
\sum_{\v k}
\Big[
-2  c^\dag_A(-\v k) \al_{\v k}  c^\dag_B(\v k) + h.c.
\Big]
\nonumber\\
&\ \ +\sum_{\v k}
(1-|\al_{\v k}|^2)
[
c^\dag_A(\v k) c_A(\v k)
+c^\dag_B(\v k) c_B(\v k)]
\nonumber\\
&\ \ +\sum_{\v k} 2 |\al_{\v k}|^2
.
\end{align}
In real space, the above Hamiltonian can be rewritten as
\begin{align}
 H&= \sum_{\v k} 2 |\al_{\v k}|^2
-\sum_{\<\v i\v j\>} ( 2u c^\dag_{\v i}c^\dag_{\v j} + h.c.)
\nonumber\\
&\ \ \ + \sum_{\v i} (1-3|u|^2)c^\dag_{\v i} c_{\v i} - \sum_{\v i,
I=1,...,6} |u|^2 c^\dag_{\v i+\v\Del_I} c_{\v i}
\end{align}
where $\{\v\Del_I\}$ are six vectors
$\v\del_1-\v\del_2$,
$-\v\del_1+\v\del_2$,
$\v\del_2-\v\del_3$,
$-\v\del_2+\v\del_3$,
$\v\del_3-\v\del_1$, and
$-\v\del_3+\v\del_1$.
Note that, if we do a particle-hole conjugation
on the B-sublattice, the above pairing Hamiltonian becomes the
following hopping Hamiltonian
\begin{align}
 H&=
-\sum_{\<\v i\v j\>} ( 2u c^\dag_{\v i}c_{\v j} + h.c.)
- \sum_{\v i, I=1,...,6} (-)^{\v i}|u|^2 c^\dag_{\v i+\v\Del_I} c_{\v i}
\nonumber\\
&\ \ \ + \sum_{\v i} (1-3|u|^2)(-)^{\v i}c^\dag_{\v i} c_{\v i}
+\text{Const.}
\end{align}
where $(-)^{\v i}=1$ if $\v i$ is in A-sublattice and
$(-)^{\v i}=-1$ if $\v i$ is in B-sublattice.

\begin{figure}
\centerline{
\includegraphics[width=3in]{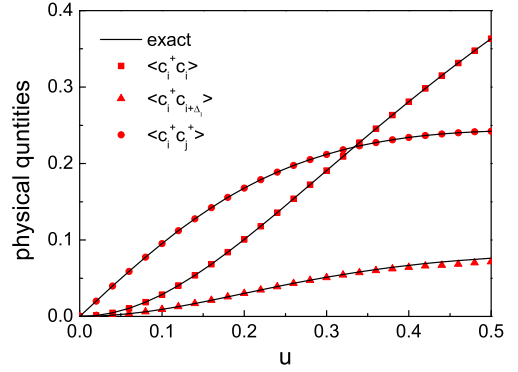}
} \caption{ \label{measurements} The expectation value of the
nearest neighbor paring, next nearest neighbor hopping and onsite
fermion number terms from the Grassmann-number tensor-entanglement
renormalization algorithm. We compare our calculation with the exact
results and find a good agreement for a large range of $u$. }
\end{figure}

\begin{figure}
\centerline{
\includegraphics[width=3in]{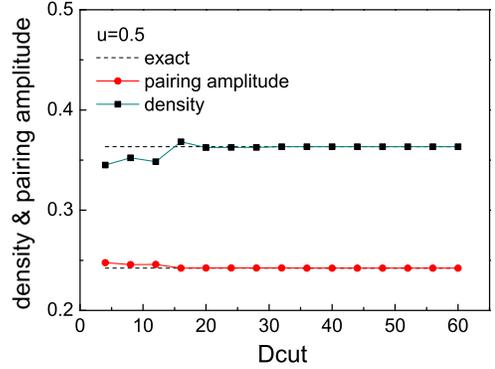}
} \caption{ \label{pairing} The expectation value of the nearest
neighbor paring and onsite fermion number at different $D_{cut}$(the
truncation dimension in the SVD decomposition) in the
Grassmann-number tensor-entanglement renormalization algorithm. It
shows all those values converge to the exact one at sufficient large
$D_{cut}$. In this calculation we choose $u=0.5$.}
\end{figure}

Fig. \ref{measurements} shows the expectation values of the nearest
paring term $c_{\v i}^\dagger c_{\v j}^\dagger$, next nearest
neighbor hopping term $c^\dag_{\v i+\v\Del_I} c_{\v i}$ as well as
onsite fermion number term $c^\dag_{\v i} c_{\v i}$ calculated from
the Grassmann-number tensor renormalization algorithm. We found very
good agreement with the exact results are found for a large range of
parameter $u$. The SVD truncation dimension is set to be
$D_{cut}=32$ throughout the whole calculation. Fig. \ref{pairing}
shows the $D_{cut}$ dependence for the physical quantities we
calculated. We find excellent agreement with exact results for large
enough $D_{cut}$.

\section{Summary}

The projective construction has played an very important role in
understanding strongly correlated systems. It can explain the
emergence of fermions, anyons (Abelian and non-Abelian), and gauge
theory in quantum spin liquids and quantum Hall states.
Mathematically, the projective construction also provides an
efficient encoding for many-body bosonic/fermionic states.

Recently, Fermionic projected entangled pair states have also been
introduced as an alternative method for efficiently encoding
many-body fermionic states. In this paper, we show that the strongly
correlated bosonic/fermionic states obtained from the projective
construction and fPEPS approach can be represented systematically as
Grassmann tensor product states.  These Grassmann tensor product
states allow us to encode many-body bosonic/fermionic states with a
number of parameters that only scales  polynomially in the number of
fermions. We have also shown that it is possible to generalize the
tensor-entanglement renormalization group (TERG) method for complex
tensor networks to Grassmann tensor networks. This allows us to
perform  an approximate calculation of the norm and average local
operators of Grassmann tensor product states in polynomial time.

In conclusion, Grassmann tensor product states can be a starting
point for a new variational approach to strongly correlated
bosonic/fermionic systems, as they not only include tensor product
state and fermion ELF states, but also systematically generalize the
slave-particle projective construction. All the physical properties
of these states can be efficiently calculated based on the Grassmann
tensor-entanglement renormalization group (GTERG) algorithm. Many
non-trivial fermionic models and frustrated spin models, such as t-J
model, Hubbard model, as well as Kagome Heisenberg model will be
studied in our future work.

We would like to thank Leon Balents, Matthew P. A. Fisher,  J.
Ignacio Cirac and Zhenghan Wang for very helpful discussions. ZCG
especially thanks the warm hospitality from Perimeter Institute for
Theoretical Physics in Canada, where this work is started. This
research is supported in part by the NSF Grant No. NSFPHY05- 51164
,the ERC grant QUERG and the FWF grant FoQuS. XGW is supported by
NSF Grant No. DMR- 0706078.

\appendix

\section{Calculate the physical measurements}

In this section, we explain how to use coarse graining
transformation to calculate the physical quantities of a Grassmann
tensor product state. As a simple example, we first explain how to
calculate the expectation value for nearest neighbor electron
pairing term $\langle \Psi|c_i^\dagger c_j^\dagger|\Psi \rangle$.
Such a term can be represented as a Grassmann tensor network with
two impurity tensors, see Eq. (\ref{pairGTPS}).

\begin{figure}
\centerline{
\includegraphics[width=3in, height=1.5in]{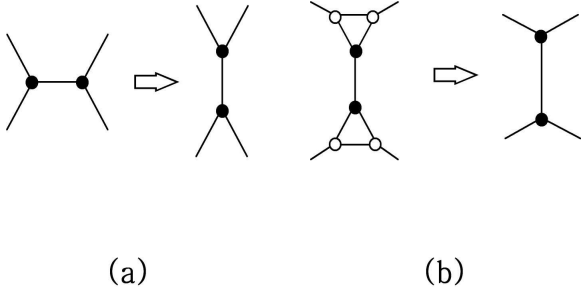}
} \caption{ \label{honeycomb1} A schematic plot of how to apply
coarse graining transformations for two nearest neighbor impurity
tensors on a honeycomb lattice. The filled dots represent impurity
tensors and open dots represent uniform tensors. In (a), we first
combine two rank three impurity tensors into a rank four impurity
tensor and then split it into two new rank three impurity tensors.
In (b), we apply the last step to combine one impurity tensor with
two uniform tensors to produce a new impurity tensor.}
\end{figure}

 We can apply the coarse
graining transformation for the uniform part in the same way as we
calculate the norm. The extra thing we need to know is how to apply
the coarse graining transformation for the two impurity tensors. It
turns out we only need to modify the first step, where we combine
the two rank three impurity tensors into a rank four impurity tensor
\begin{equation}
\label{imTTTG}
 \v T_{p_1p_2p_3p_4}^{\prime\prime\prime}=\sum_{p_5p_6} \int_{ij}
\v T_{i;p_5p_1p_2}^\prime \v T_{j;p_6p_3p_4}^{\prime\prime} \v
G_{ij;p_5p_6},
\end{equation}
with
\begin{widetext}
\begin{align}
\label{imexpTG} & \v T_{i;p_5p_1p_2}^\prime=
\sum_{\{n^{\al_5}_5\}\{n^{\al_1}_1\} \{n^{\al_2}_2\}}
{\cT^\prime}^{\{n^{\al_5}_5\}\{n^{\al_1}_1\}\{n^{\al_2}_2\}}_{i;p_5p_1p_2}
\widetilde{\prod_{\al_5}} (\dd \eta^{\al_5}_{5})^{n^{\al_5}_5}
\widetilde{\prod_{\al_1}} (\dd \eta^{\al_1}_{1})^{n^{\al_1}_1}
\widetilde{\prod_{\al_2}} (\dd \eta^{\al_2}_{2})^{n^{\al_2}_2}
\nonumber\\
& \v T_{j;p_6p_3p_4}^{\prime\prime}=
\sum_{\{n^{\al_6}_6\}\{n^{\al_3}_3\} \{n^{\al_4}_4\}}
{\cT^{\prime\prime}}^{\{n^{\al_6}_6\}\{n^{\al_3}_3\}\{n^{\al_4}_4\}}_{j;p_6p_3p_4}
\widetilde{\prod_{\al_6}} (\dd \eta^{\al_6}_{6})^{n^{\al_6}_6}
\widetilde{\prod_{\al_3}} (\dd \eta^{\al_3}_{3})^{n^{\al_3}_3}
\widetilde{\prod_{\al_4}} (\dd \eta^{\al_4}_{4})^{n^{\al_4}_4}
\nonumber\\
& \v T_{p_1p_2p_3p_4}^{\prime\prime\prime}=
\sum_{\{n^{\al_1}_1\}\{n^{\al_2}_2\}\{n^{\al_3}_3\} \{n^{\al_4}_4\}}
{\cT^{\prime\prime\prime}}^{\{n^{\al_2}_2\}\{n^{\al_2}_2\}\{n^{\al_3}_3\}\{n^{\al_4}_4\}}
_{p_1p_2p_3p_4} \widetilde{\prod_{\al_1}} (\dd
\eta^{\al_1}_{1})^{n^{\al_1}_1} \widetilde{\prod_{\al_2}} (\dd
\eta^{\al_2}_{2})^{n^{\al_2}_2} \widetilde{\prod_{\al_3}} (\dd
\eta^{\al_3}_{3})^{n^{\al_3}_3} \widetilde{\prod_{\al_4}} (\dd
\eta^{\al_4}_{4})^{n^{\al_4}_4}
\nonumber\\
& \v G_{ij;p_5 p_6}=\sum_{\{n^{\al_5}_5\}\{n^{\al_6}_6\}}
\cG^{\{n^{\al_5}_5\}\{n^{\al_6}_6\}}_{ij;p_5 p_6} \prod_{\al_5}
(\eta^{\al_5}_{5})^{n^{\al_5}_5} \prod_{\al_6}
(\eta^{\al_6}_{6})^{n^{\al_6}_6}.
\end{align}
Because $\v T^\prime$ and $\v T^{\prime\prime}$ contain an odd
number of Grassmann numbers in this case, the relationship of the
coefficients should be modified as:
\begin{align}
{\cT^{\prime\prime\prime}}^{\{n^{\al_2}_2\}\{n^{\al_2}_2\}\{n^{\al_3}_3\}\{n^{\al_4}_4\}}
_{p_1p_2p_3p_4} = \sum_{p_5p_6}
\sum_{\{n^{\al_5}_5\}\{n^{\al_6}_6\}} (-)^{ (\sum_{\al_5}
n^{\al_5}_5)}
{\cT^\prime}^{\{n^{\al_5}_5\}\{n^{\al_1}_1\}\{n^{\al_2}_2\}}_{i;p_5p_1p_2}
\cG^{\{n^{\al_5}_5\}\{n^{\al_6}_6\}}_{ij;p_5 p_6}
{\cT^{\prime\prime}}^{\{n^{\al_6}_6\}\{n^{\al_3}_3\}\{n^{\al_4}_4\}}_{j;p_6p_3p_4}.
\end{align}
\end{widetext}
The second step is the same as we calculate the norm because the
rank four impurity tensor contain an \emph{even} number of Grassmann
numbers. Of course the last step also remains the same except we
need to combine one new impurity tensor with two new uniform tensors
to produce a coarse grained impurity tensor, as we do in the
standard TERG algorithm, see in Fig. \ref{honeycomb1}. If we
calculate the expectation value for two nearest neighbor bosonic
operators, such as $\<n_in_j\>$, even the first step dose not need
to be modified because the impurity tensors contain even number of
Grassmann numbers in this case.

\begin{figure}
\centerline{
\includegraphics[width=2.5in]{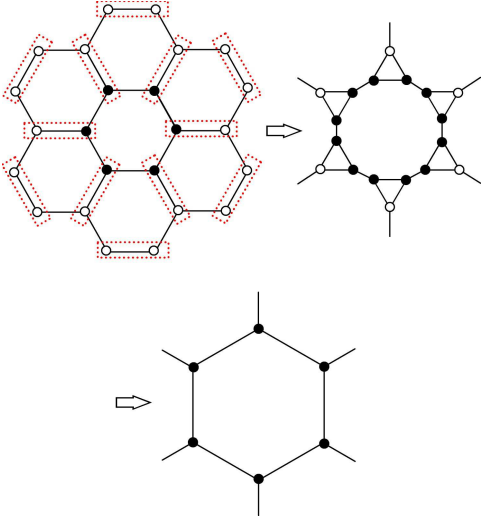}
} \caption{ \label{honeycomb2} A schematic plot of the coarse
graining transformations for generic six body interactions on the
hexagon of a honeycomb lattice. The filled dots represent impurity
tensors and open dots represent uniform tensors.}
\end{figure}

For generic interactions as well as correlation functions, the
coarse graining procedure can be designed in the same way as in the
TERG algorithm. Fig. \ref{honeycomb2} shows how to do the coarse
graining transformation for generic six body interactions on the
hexagon of a honeycomb lattice. However, we need to take care of the
sign factor when we apply the coarse graining transformation for
impurity tensors which contain an odd number of Grassmann number and
in the most generic case, all the three steps need to be modified.
For example, in the second step, if the rank four impurity tensor
$\v T^{\prime\prime\prime}$ contain an odd number of Grassmann
numbers, then we need to decompose it into two new rank three
impurity tensors where one has an odd number of Grassmann numbers
and the other has an even number of Grassmann numbers. We also need
to take care of the sign factor in the last step if we combine
impurity tensors with odd number of Grassmann numbers.

\section{A review of projective construction for fractional quantum
Hall state}
\label{pcon}

In this section, we are going to review two examples of fractional
quantum Hall state to demonstrate how to obtain low energy effective
theory from a projective approach.\cite{Wnab,Wpcon}

To use the projective construction to study the $\nu=1/3$ Laughlin state
of $N$ electrons, we consider a system of three kinds of partons
described by the fermion operators $\psi_a(z)$, $a=1,2,3$.  The parton
system contain $N$ particles for each kind of partons, and each kind
of parton form $\nu=1$  state (which is a ELF state).  Let us
denote the $\nu=1$  state for the $a^\text{th}$ partons as
$|\Psi_1\>_a$. Then the total ground state of the system of three
kinds of partons is given by $ |\Psi_1\>_1\otimes |\Psi_1\>_2\otimes
|\Psi_1\>_3 \equiv |\Psi_1\Psi_1\Psi_1\> $.  Note that the state
$|\Psi_1\Psi_1\Psi_1\>$ is a ELF state.  Then the
$\nu=1/3$ Laughlin wave function can be expressed as a projection of the
ELF state $|\Psi_1\Psi_1\Psi_1\> $:
\begin{align}
\label{Psi3proj}
 \Psi_3(\{z_i\})=
\<0|\prod_i [
\psi_1(z_i)
\psi_2(z_i)
\psi_3(z_i)]|\Psi_1\Psi_1\Psi_1\>.
\end{align}
Note that the projection simply combine the three partons into a single
electron:
\begin{equation}
c(z)= \psi_1(z) \psi_2(z) \psi_3(z) .
\end{equation}
Although the expression \eq{Psi3proj} is very formal, the projection
that it describes can be done at the field theory level, which
allows us to calculate the low energy effective theory of the
Laughlin state.

To do the projection at the field theory level, we start with the
Lagrangian that describes three independent partons (that is before
the projection)
\begin{align}
 \cL_0 = \sum_{a=1}^3
\Big[
\psi_a^\dag \imth\prt_t \psi_a
-\frac{1}{2m}
|(
\boldsymbol{\prt}
-\imth \frac{e}{3} \v A) \psi_a|^2
-\mu \psi_a^\dag \psi_a
\Big],
\end{align}
where we have assumed that each kind of parton carries $e/3$
electric charge and $\v A$ is the vector potential that describes
the uniform magnetic field.  The chemical potential is chosen such
that each kind of parton fills its first Landau level and forms the
$\nu=1$ quantum Hall state.

We note that the above Lagrangian that describes the independent
partons before the projection has an $SU(3)$ symmetry:
\begin{equation}
 \psi_a \to U_{ab}\psi_b,\ \ \ U \in SU(3).
\end{equation}
The theory of independent partons contains fluctuations of the density
and the current of the $SU(3)$ charge.  On the other hand, we see that
the electron operator $c$ transforms as
\begin{equation}
c= \psi_1 \psi_2 \psi_3 \to \det(U) \psi_1 \psi_2 \psi_3 = c
\end{equation}
since the $\psi_a$ anticommute with each other.  Thus the electron
operator is invariant under the $SU(3)$ transformation. As a result,
the electronic state obtained after the projection in \eq{Psi3proj}
is also invariant under the $SU(3)$ transformation since both $|0\>$
and $c$ are $SU(3)$ invariant.  This means that, after the
projection, the electronic state contains no fluctuations of the
density and the current of the $SU(3)$ charges.

This motivates us to perform the projection at the field theory
level by including a $SU(3)$ gauge theory in the above
independent-parton model:
\begin{align}
 \cL_p &=
\sum_{a,b}
\psi_a^\dag \imth[\del_{ab}\prt_t-\imth (a_0)_{ab}] \psi_b
\\
&\ \ \ \
+\frac{1}{2m}
\sum_{a,b}
\psi^\dag_a
( \boldsymbol{\prt} -\imth \frac{e}{3} \v A-\imth \v a)^2_{ab} \psi_b
-\mu \sum_a \psi_a^\dag \psi_a ,
\nonumber
\end{align}
where $(a_0, a_x, a_y)$ are the $SU(3)$ gauge fields which are 3 by 3
hermitian matrix valued fields.
The $SU(3)$ gauge fields remove all
the $SU(3)$ density and current fluctuations.
In other word, if we perform the
path integral of the $SU(3)$ gauge fields first
\begin{align}
 \e^{-\imth \int dtd^2\v x\;\cL_e}=\int D[\v a]D[a_0]
\e^{-\imth \int dtd^2\v x\; \cL_p},
\end{align}
the resulting effective theory $\cL_e$ will contain no $SU(3)$
fluctuations.  Thus we say that the path integral of the $SU(3)$ gauge
fields, $\int D[\v a]D[a_0]$, performs the projection at the field theory
level.

The full theory is described by the path integral
over both parton fields $\psi_a$ and $SU(3)$ gauge fields $(a_0, a_x, a_y)$:
\begin{align}
 Z=\int D[\psi_a] \int D[\v a]D[a_0] \e^{-\imth \int dtd^2\v x\; \cL_p}
\end{align}
If we exchange the integration order:
\begin{align}
 Z &=
\int D[\v a]D[a_0]
\int D[\psi_a]
\e^{-\imth \int dtd^2\v x\; \cL_p}
\nonumber\\
&=
\int D[\v a]D[a_0]
\e^{-\imth \int dtd^2\v x\; \cL_a},
\end{align}
we obtain an effective theory that contain only the $SU(3)$ gauge
field $\cL_a(a_0,a_x,a_y)$.  Since each kind of parton forms the
$\nu=1$ quantum Hall state, the resulting $SU(3)$ effective theory
turns out to be the level-1 $SU(3)$ Chern-Simons theory.  We
conclude that the $\nu=1/3$ Laughlin state is described by the
level-1 $SU(3)$ Chern-Simons topological field theory.  All the
topological properties of the $\nu=1/3$ Laughlin state, such as
fractional charges and fractional statistics can be obtained from
such a $SU(3)_1$ Chern-Simons theory.  Note that the level-1 $SU(3)$
Chern-Simons theory is equivalent to a $U(1)$ Chern-Simons theory.
Thus $SU(3)_1$ Chern-Simons theory actually describes an Abelian
state.

We have seen that if we let partons to form the $\nu=1$ quantum Hall
state, we will obtain the $\nu=1/3$ Laughlin state and the $SU(3)_1$
Chern-Simons  theory as its low energy effective theory.  If we let
each of the three kinds of partons forms the $\nu=m$ quantum Hall
state, we will obtain a $\nu=m/3$ quantum Hall state and the
level-$m$ $SU(3)$ Chern-Simons theory as its low energy effective
theory.\cite{Wnab} Such a $\nu=m/3$ quantum Hall state is a
non-Abelian quantum Hall state.

Similarly, starting from the ELF state for four kinds
of partons $|\Psi_1\Psi_1\Psi_1\Psi_1\> $, we can construct the
non-Abelian $\nu=1$ Pfaffian state (for bosonic electrons)
\begin{align}
\label{Pafproj}
&\ \ \ \Psi_\text{Pfa}(\{z_i\})
\\
&= \<0|\prod_i [ \psi_1(z_i) \psi_4(z_i)
-\psi_3(z_i)\psi_2(z_i)]|\Psi_1\Psi_1\Psi_1\Psi_1\> .
\nonumber
\end{align}
where the electron operator is related to the parton operators as
\begin{equation}
c(z)=\psi_1(z) \psi_4(z) -\psi_3(z)\psi_2(z).
\end{equation}

To do the projection at the field theory level, we start with the
independent parton model
\begin{align}
 \cL_0 = \sum_{a=1}^4
\Big[
\psi_a^\dag \imth\prt_t \psi_a
-\frac{1}{2m}
|(
\boldsymbol{\prt}
-\imth \frac{e}{2} \v A) \psi_a|^2
-\mu \psi_a^\dag \psi_a
\Big],
\end{align}
We note that the above Lagrangian
has a $SU(4)$ symmetry:
\begin{equation}
 \psi_a \to U_{ab}\psi_b,\ \ \ U \in SU(4).
\end{equation}
Thus the theory of independent partons contains $SU(4)$ charge and
current fluctuations.

However, the electron operator $c=\psi_1 \psi_4 -\psi_3\psi_2$ is not
invariant under the full $SU(4)$ transformations.  If we identify
$(\psi_1,...,\psi_4)=(\psi_{11},\psi_{12},\psi_{21},\psi_{22})$, we
find
\begin{align}
 c &=\psi_{11} \psi_{22} -\psi_{21} \psi_{12}
\\
 &\propto
\psi_{11} \psi_{22}
- \psi_{21} \psi_{12}
+ \psi_{12} \psi_{21}
- \psi_{22} \psi_{11}
\nonumber\\
&\propto
\psi_{a\al}
(\tau_2)_{ab} (\si_1)_{\al\bt}
\psi_{b\bt}
\propto \psi^T
(\tau_2\otimes \si_2)
(\tau_0\otimes \si_3)
\psi  ,
\nonumber
\end{align}
where $\tau_0=\si_0$ are the 2 by 2 identity matrix and $\tau_l$,
$\si_l$ are the Pauli matrices.  Here $\tau_l$ acts on the first
subscript $a$ of $\psi_{a\al}$ while $\si_l$ acts on the second
subscript $\al$ of $\psi_{a\al}$.  We see that the electron operator
is invariant under a subgroup of $SU(4)$ generated by 10 generators:
$\tau_i \otimes \si_0$, $\tau_i\otimes\si_1$, $\tau_i\otimes\si_2$,
and $\tau_0 \otimes \si_3$, It turns out that the above 10
generators generate the $SO(5)$  group in its 4 dimensional spinor
representation.  Therefore, the electronic states do not contain any
$SO(5)$ fluctuations.

To remove the $SO(5)$ fluctuations at the field theory level, we can
include a $SO(5)$ gauge field (the 4 dimensional representation
spanned by $\tau_i \otimes \si_0$, $\tau_i\otimes\si_1$,
$\tau_i\otimes\si_2$, and $\tau_0 \otimes \si_3$) in the parton
Lagrangian.  After integrating out the partons we obtain the low
energy effective Chern-Simons theory for the Pfaffian state, which
is a $SO(5)$ Chern-Simons theory.  A different effective
Chern-Simons theory for the Pfaffian state was obtained in
\Ref{FNT9804}.

We would like to point out that in the first form of the projective
construction \eq{Psi3proj}, the electron operator $c(z)$ is
expressed as a product of parton operators.  In this case, we can
use variational Monte Carlo calculation to numerically study many
properties (such as ground state energy) of such projective states.
On the other hand, in the second form of the projective construction
\eq{Pafproj}, the electron operator $c(z)$ is expressed as a sum of
several products of parton operators.  In this case, in general, the
variational Monte Carlo method is ineffective due to the sign
problem. So far, we still do not have an effective numerical method
for the second form of the projective construction.

The Grassmann tensor network can represent projective states
obtained from the both forms of the projective construction.  So the
renormalization of the Grassmann tensor network might allow us to
approximately calculate the norms and average local operators for
both forms of projective states. A detailed study will be presented
in our future work.


\end{document}